# Crystal Chemical Concept of Arrangement and Function of Layered Superconducting Materials

L.M. Volkova[1], S.A. Polyshchuk[1], S.A. Magarill[2], and F.E. Herbeck[3]

**ABSTRACT**

The crystal chemical concept of arrangement and function of layered superconducting materials is supposed. The concept is based on results of our investigation of crystal chemistry of high-temperature superconductors (HTSC) cuprates, diborides $AB_2$ and borocarbides of nickel $RNi_2B_2C$. According to these results: (1) the main role in appearance of superconductivity played by the structural fragments – sandwiches $A_2(CuO_2)$ in HTSC cuprates, $A_2(B_2)$ in diborides and $RB(Ni)$ in nickel borocarbides but not the separate planes of $CuO_2$, $B_2$ or Ni; (2) correlations between $T_c$ and crystal chemical parameters of these sandwiches have similar character in all three classes of compounds, despite of distinction of a nature of their superconductivity. The central idea of the concept consists in following: in contrast to metallic conduction, for which it is enough to provide only concentration and mobility of charge carriers, for occurrence of a superconductivity it is necessary to create in addition a space (channels) for stream of charge carriers, compression of stream of carriers and focusing to direction on path of motion.

## 1. INTRODUCTION

The complexity of the phenomenon defines a state of researches in the field of superconductivity. It concerns also to investigation a crystal chemistry of high- and low-temperature superconductors. For 16 years of research of high-temperature superconductors (HTSC) every possible family of complex layered oxides of copper were synthesized, and at a high experimental level both neutron and X-ray diffraction studies and definitions of superconducting properties of these compounds are carried out. In result the variation of structure and superconducting transition temperature ($T_c$) under effect of pressure and at different substitutions of cations or anions was investigated. For many compounds the position of each atom in a crystal before and after transition in a superconducting state was established. At the same time, it was not possible to understand wholly the phenomenon of superconductivity. After discovery HTSC in cuprates [1] the search and prediction of new superconducting compounds was intensively carried on. The recent opening of superconductivity in $MgB_2$ [2] became the stimulus for further development of investigations in this direction. On the one hand, such search is especially complex without the unique theory explaining the mechanism HTSC, and with another, the practice had shown, that the microscopic theories of a superconductivity, for example BCS theory, do not answer a question, what substances can be and what cannot be superconductors. As for now, the prediction of superconductors most successfully can be fulfilled only on a basis of the crystal chemical approach, bound with investigation of correlations between composition, structure and transport properties of compounds, with engaging results of various theoretical calculations.

[1] Institute of Chemistry, Far Eastern Branch of Russian Academy of Science, 690022 Vladivostok, Russia
[2] Institute of Inorganic Chemistry, Sib. Branch of Russian Academy of Science, 630090 Novosibirsk, Russia
[3] Institute of Automation and Control Processes, Far Eastern Branch of Russian Academy of Science, 690041 Vladivostok, Russia.

## 2. BASE OF CONCEPT: SIMILARITY OF SUPERCONDUCTING FRAGMENTS AND CRYSTAL CHEMICAL CORRELATIONS IN CUPRATES, DIBORIDES AND NICKEL BOROCARBIDES

It is common practice to connect the temperature of transition in a superconducting state, like the majority of properties HTSC cuprates, predominantly with concentration of charge carriers and, accordingly, with optimal interatomic distances in a CuO2 plane, containing these carriers, and with number of CuO2 planes in a superconductor. These positions are the basic both to simulation of new superconductors, and to construction of various crystal chemical dependences. However dependences, obtained on this basis, are not universal and are fulfilled only in limits of several families of HTSC. We have shown [3-6], that for occurrence of a superconductivity it is fundamental importance not only have a separate plane with charge carriers, but also the extended structural fragment, which arrangement and function allows to have stream of charge carriers and besides the space for its carry and resource of effect on this stream.

On the basis of an extensive experimental literary material on layered superconductors with various nature of superconductivity, we found at first the universal for all phases of HTSP cuprates formed by single $CuO_2$-plane and by several $CuO_2$-planes empirical correlations between $T_c$ and crystal chemical parameters of a cation sublattice $A_2(Cu)$ of sandwich $A_2(CuO_2)$ [3-4] (Fig. 1 a). We have proved by this, that a structural fragment, where a superconductivity can be, is not only one plane of $CuO_2$, but sandwich of $A_2(CuO_2)$, and have shown, that at presence in a superconductor of several such fragments, for example, in HTSC cuprates with several CuO2 planes, the value of the $T_c$ of compound is defined by intrinsic $T_c$ of that fragment, which have more high $T_c$ (whose parameters are closer to optimal). It was recently confirmed by experimental [7] that interaction between $CuO_2$ planes is not critical for superconductivity of HTC cuprates.

Then [5, 6] we have installed, that in layered quasi-two-dimensional systems, such as diborides $AB_2$ and borocarbides of nickel $RNi_2B_2C$ (R = rare earth), it is possible to select the elementary structural fragments of $A_2(B_2)$ and RBNi, similar to fragments of $A_2(Cu)$, existing in HTSC cuprates (Fig. 1b and c). These fragments represent the sandwiches I/S/I, where the internal plane S contains charge carriers capable to carry, and the external planes I are dielectric ones. Correlations between $T_c$ and relation ($J$) [3-6] values of crystal chemical parameters of these sandwiches have similar character in all three classes of compounds, despite of distinction of a nature of their superconducting properties, and are represented by parabolic curves (Fig. 2). The value $J$ is determined as a quotient of distances $d$(M-M) between atoms of copper, boron or nickel located in an internal plane sandwich by the total of effective distances $(D_1 + D_2)$ between internal and external planes sandwich, taking in to account by calculation a charge and size of ions forming these planes, and also difference in these values at substitutions ($J = d(M-M)/(D_1 + D_2)$). The concentration of charge carriers is taken into account in correlations through the mediation of structural parameters of sandwich.

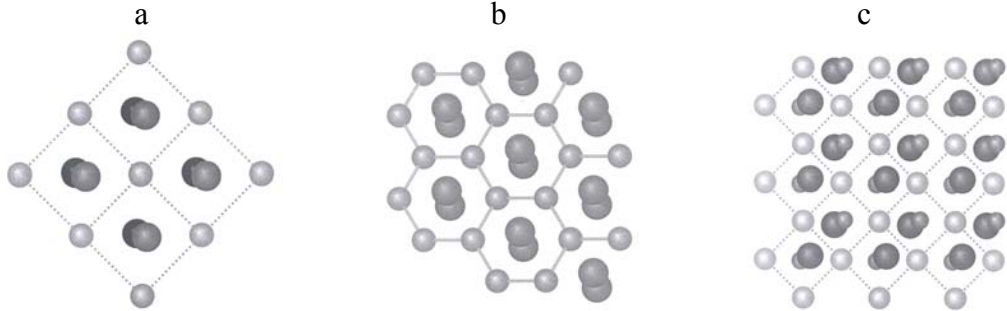

**Fig. 1**. Structural fragments – sandwiches: ); $A_2(Cu)$ in HTSC cuprates (A – large ball, Cu – small ball (a); $A_2(B_2)$ in diborides (A-large ball, B small ball) (b); and RB(Ni) in $RNi_2B_2C$ (R – large black ball, B – small black ball, Ni – light ball) (c).

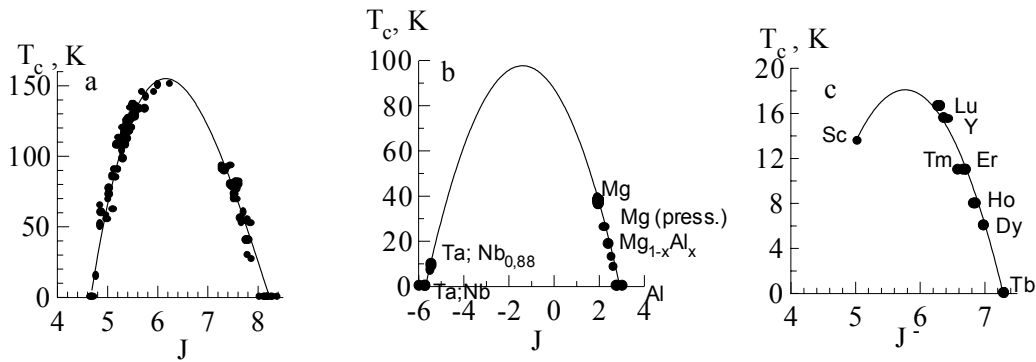

**Fig. 2.** Correlation $T_c$ with a relation ($J$ and $J$) of crystal chemical parameters in HTSC cuprates with several CuO2 planes [4] (a); in diborides $AB_2$ [5] (b); and in borocarbides $RNi_2B_2C$ [6] (c).

High $T_c$ in doping by holes cuprates and diboride $MgB_2$ in comparison with $T_c$ in doping by electrons cuprates and in borocarbides $RNi_2B_2C$ it is accepted to explain by a nature of charge carriers. However, if to consider as superconducting fragment not a plane, but sandwich, it is possible to explain this experimental fact differently: a the reaching of high temperatures of transition is apparently bound with more effective focusing action on carriers (holes) of a field of like charged ions (cations) located in external planes of sandwich. In addition, the features of a structure borocarbides of nickel have allowed us to show [6], that bonding electrons (between atoms Ni and B), located on trajectory of charge carriers transfer, can suppress a superconductivity.

## 3. CRYSTAL CHEMICAL CONCEPT OF ARRANGEMENT AND FUNCTION OF LAYERED SUPERCONDUCTING MATERIALS

As a result of these researches the concept of arrangement and function of layered superconducting materials were formulated. The main idea of this concept consists is the following: in contrast to metallic conduction, for which it is enough to provide only concentration and mobility of charge carriers, for occurrence of a superconductivity it is necessary to create in addition a space (channels) for stream of charge carriers, compression of stream of carriers and focusing to a direction on path of motion. There are strong grounds to believe that the elementary structural fragment, consisting of a core and the shell, is responsible for occurrence of superconductivity in compound. The core should create a continuous stream of charge carriers, and be enclosed in the dielectric shell formed by ions (more preferably like charged with carriers). The function of the shell (extended ionic collimator) consists in shaping of this stream: compression and focusing of stream of charge carriers to a direction of motion. Such arrangement, probably, results in an effect reminding a planar channelling of charged particles, and in this case of stream of pairs of charged particles.

The critical parameters defining a possibility of occurrence of a superconductivity and the value of the $T_c$, are concentration of charge carriers, distance between atoms forming a core, parameters of a transversal size of space between a core and shell, and also charge and degree of a homogeneity of the shell. Presence of defects in a core, overlap of space of charge carriers, and also unlikeness of charges of carriers and the shell can result in to lowering of the $T_c$ down to suppression of a superconductivity.

We suppose that in one-dimensional and three-dimensional superconductors it would be also possible to select extended one- or two-dimensional fragments fulfilling similar functions, to find correlation of the $T_c$ with crystal chemical parameters of these fragments and by that to expand this concept. It is probably that the core of a superconducting fragment can represent not only simple net or linear chain of atoms, but also cluster formation (for example, in $Tl_2Mo_6Se_6$ ($T_c$ = 6.6 K) [8] condensed in chain $Mo_6$ octahedrons in shell from Se). The conception can be utilised for search and modelling of new superconductors, including high-temperature.

## ACKNOWLEDGMENTS

The work was supported by RFFI (Russia) (grant № 00-03-32486).